\documentclass[iop,apj,numberedappendix,appendixfloats]{emulateapj}

\usepackage{natbib}
\usepackage{longtable}
\usepackage{rotating}
\usepackage[usenames,dvipsnames]{color}
\usepackage{enumerate}
\usepackage{amsmath}

\usepackage[colorlinks=true,citecolor=blue,linkcolor=blue]{hyperref}

\usepackage{CJK}

\def\arcsec{\hbox{$^{\prime\prime}$}}
\def\arcmin{\hbox{$^{\prime}$}}

\def\g0{G$_{0}$}
\def\2cm{cm$^{-2}$}
\def\cm3{cm$^{-3}$}
\def\kms{km s$^{-1}$}

\def\nh3{NH$_3$}
\def\n2h{N$_2$H$^+$}
\def\C2h{C$_2$H}
\def\co{$^{12}$CO}
\def\13co{$^{13}$CO}
\def\c18o{C$^{18}$O}
\def\hc3n{HC$_3$N}
\def\h2{H$_2$}
\def\nh{n(H$_2$)}

\def\chp{CH$^{+}$}
\def\ch3p{CH$_{3}^{+}$}

\newcommand{\cii}{C\,{\sc ii}}
\def\lp{\>\> .}
\def\lc{\>\> ,}

\def\arcsec{\hbox{$^{\prime\prime}$}}
\def\arcmin{\hbox{$^{\prime}$}}

\shorttitle{}
\shortauthors{}
\slugcomment{}

\begin{document}
\begin{CJK*}{UTF8}{gbsn}    


\title{CH as a Molecular Gas Tracer and C-Shock Tracer Across a Molecular Cloud Boundary in Taurus}

\author{Duo Xu (许铎) \altaffilmark{1,2,4}{*}, Di Li (\CJKfamily{bsmi}李菂) \altaffilmark{1,3}{*} }
\affil{
$^1$ National Astronomical Observatories, Chinese Academy of Sciences, A20 Datun Road, Chaoyang District, Beijing 100012, China\\
{$^2$ University of Chinese Academy of Sciences, Beijing 100049, China\\} 
$^3$ Key Laboratory for Radio Astronomy, Chinese Academy of Sciences, China\\
$^4$ Current address: Department of Astronomy, University of Massachusetts, Amherst, MA 01003, USA
}

\email{{*}Email: dxu@astro.umass.edu, dili@nao.cas.cn}

\begin{abstract}
We present new observations of all three ground-state transitions of the methylidyne (CH) radical and all four ground-state transitions of the hydroxyl (OH) radical toward a sharp boundary region of the Taurus molecular cloud. These data were analyzed in conjunction with existing CO and dust images. The derived CH abundance is consistent with previous observations of translucent clouds ($0.8\le A_{v} \le 2.1$ mag). The $X({\rm CH})$-factor is nearly a constant at $(1.0\pm0.06)\times10^{22}$ $\rm {cm^{-2}~K^{-1}~km^{-1}~s}$ in this extinction range, with less dispersion than that of the more widely used molecular tracers CO and OH. CH turns out be a better tracer of total column density in {such an} intermediate extinction range than CO or OH. Compared with previous observations, CH is overabundant below 1 mag extinction. Such an overabundance of CH is consistent with the presence of a C-shock. CH has two kinematic components, one of which shifts from 5.3 to 6 \kms, while the other stays at 6.8 \kms\ when moving from outside toward inside of the cloud. These velocity behaviors exactly match with previous OH observation. The shifting of the two kinematic components indicates colliding streams or gas flow at the boundary region, which could be the cause of the C-shock.

\end{abstract}


\section{Introduction}

The 3.3 GHz $\Lambda$-doubling lines of the methylidyne radical CH have been commonly used to trace low density gas in diffuse clouds and at the boundaries of dense clouds \citep[e.g.][]{1989ApJ...339..244M, 1993ApJ...408..559M}. They have also been observed extensively toward translucent clouds, bright limbed clouds, outflows, and dark clouds \citep[e.g.][]{1978ApJ...224..125L, 1980A&A....83..226S, 1981A&A....97..317S, 1986A&A...160..157M, 1987A&A...173..347J, 1988ApJ...329..920S, 1992A&AS...93..509M}. Many of such surveys have shown a linear correlation between the CH column density and the visual extinction in diffuse and translucent clouds \citep[e.g.][]{1977ApJS...35..263H, 1986A&A...160..157M, 1993ApJ...408..559M, 2005AJ....130.2725M}. CH is therefore recognized as a powerful tracer of molecular hydrogen, tightly correlated with molecular hydrogen in terms of column density $\rm [CH]/[H_{2}] = 3.5\times10^{-8}$ \citep{2008ApJ...687.1075S}. {However, in a high spatial resolution observation of two high-latitude translucent clouds MBM 3 and 40 with the Arecibo Telescope, \citet{2010AJ....139..267C} found a slight spatial offset between the distribution of CH and that of CO, possibly due to the  chemical evolution of carbon.} At higher visual extinctions, CH is consumed by the carbon chemistry in relatively dense molecular environments \citep{1986A&A...160..157M, 1986A&AS...64..391V}. Thus, CH observations cannot supplant CO for studying {high-visual-extinction} clouds such as dark clouds or giant molecular clouds.

Owing to different formation routes, CH can exist in both high and low density gas. In high density gas, CH can be formed through the following chains of reactions \citep{1973ApL....15...79B}:
\begin{eqnarray}
{\rm C^{+} + H_2 \rightarrow  CH_{2}^{+}\, +\, } h\nu \nonumber \\
{\rm CH^{+}_{2} + H_2 \rightarrow  CH_{3}^{+} + H} \nonumber \\
{\rm CH^{+}_{2}} + e^{-} \rightarrow {\rm CH + H} \nonumber  \\
{\rm CH^{+}_{3}} + e^{-} \rightarrow {\rm CH + H_2}  \lp
\label{chreact}
\end{eqnarray}
When the amount of ionized carbon ([\cii]) is substantial, CH formation is believed to be triggered by the radiative synthesis of [\cii] with vibrationally excited molecular hydrogen \h2\ (as described in Equation~(\ref{chreact})) in the outer layers of photodissociation regions (PDRs), where the chemical evolution is dominated by UV radiation. {Observationally, the evidence for the above equations has been inconclusive \citep[see e.g.~the \ch3p\ observation toward Cyg OB2 by][]{2010ApJ...724.1357I}.} In lower density material (${n_{H}}\sim 50$ \cm3), CH can also be produced {through} \chp\ synthesis,
\begin{eqnarray}
{\rm C^{+} + H} &\rightarrow&  {\rm CH^{+} +\, } h\nu \nonumber \\
{\rm C^{+} + H_{2} + 0.396\, eV} &\rightarrow&  {\rm CH^{+}+ H } \nonumber \\
{\rm CH^{+}} + e^{-} &\rightarrow&  {\rm CH\, +\, } h\nu  \lc
\label{chreact2}
\end{eqnarray}
{propelled by} MHD shocks \citep{1986ApJ...306..655D, 1986MNRAS.220..801P}. CH is thus considered to be a tracer of MHD shocks, especially the C-type shocks \citep{1998MNRAS.297.1182F}. \citet{1986ApJ...310..392D} made theoretical calculations on the abundance of CH when a C-type shock propagates into a diffuse cloud with $n_{\rm H} = 50$ \cm3, which may help us identify the presence of C-shocks. {However, \citet{1993A&A...269..477G} and \citet{1995MNRAS.277..458C} observed optical CH and \chp\ lines in several stars but found their line profiles to be inconsistent with shock theories. They believed that turbulent chemistry gives rise to the overabundance of CH and \chp\ in these clouds. \citet{2016ApJ...829...15M} found UV irradiation rather than shock chemistry playing a key role in \chp\ formation in Orion BN/KL. }

The boundary of molecular clouds is the region {in which} C-shock may take place \citep{xu2016}. {A clear example of cloud boundaries can be found in Taurus \citep{Goldsmith2008}, northeast of the TMC1 region with a visual extinction ranging from 0.4 to 2.7 mag, and is thus an  ideal target to study CH as a molecular tracer in the transition zone between diffuse and dense gas.} \citet{xu2016} have found evidence of the existence of C-shock across the Taurus boundary, such as the overabundance of OH at visual extinctions at or below 1 mag,  the conjugate emission of OH 1612 and 1720 MHz components, and the colliding streams or gas flow at the boundary region. All the evidence of C-shock across the Taurus boundary in \citet{xu2016} is derived from OH spectra. Further spectral analysis of CH can provide insights into the presence of C-shock across the Taurus boundary.

\begin{figure}[htp]
\centering
\includegraphics[width=1.0\linewidth]{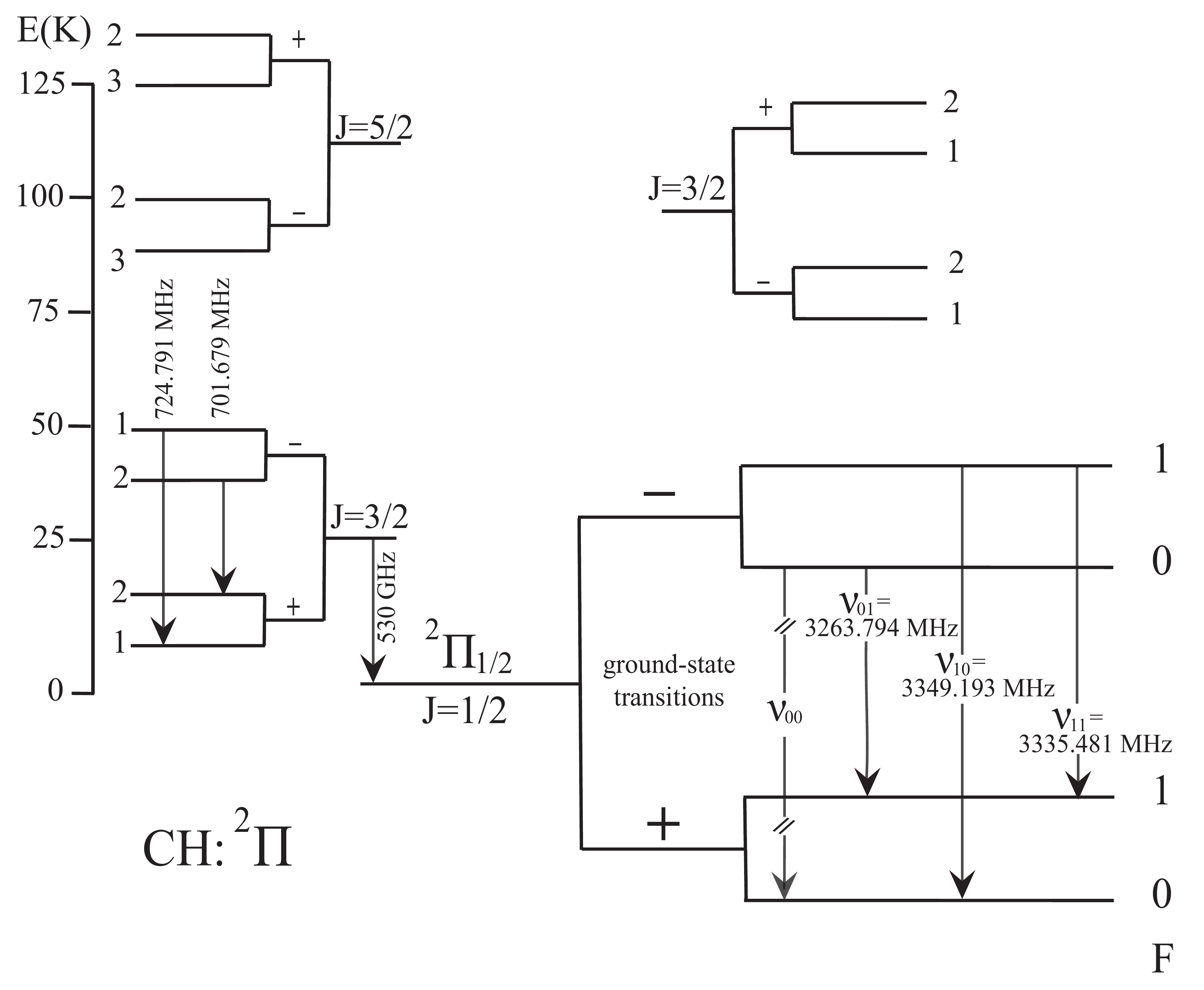}
\caption{The energy levels of CH (not shown to scale). The intensity ratio between the allowed ground state transitions is $I(\nu_{11})$:$I(\nu_{10})$:$I(\nu_{01})$ = 2:1:1 under LTE assumption.}
\label{fig.ch_energy_level_ground}
\end{figure}

We have carried out observations of the Taurus boundary in three CH transitions (3335, 3264, and 3349 MHz, as shown in Figure~\ref{fig.ch_energy_level_ground}) using the 305 m Arecibo Telescope. We made a total of five cuts {1.5 arcminutes apart} across the boundary region each with 17 pointings (Figure~\ref{fig.cut_ra_dec}). We describe the observations of CH and OH across the boundary region and the \co\ $J = 1-0$, and \13co\ $J = 1-0$ map of the Taurus molecular cloud in Section~\ref{Observations and Data}. We analyze the CH spectrum and derive CH column density across the boundary in Section~\ref{Analysis}. We discuss the advantage of CH as a molecular tracer in Section~\ref{CH As a Molecular Gas Tracer}. We compare the CH abundance with the C-shock model in Section~\ref{CH Abundance and C-shock Model}. In Section~\ref{Summary and Conclusions} we summarize our results and conclusions from this study.

\section{Observations and Data}
\label{Observations and Data}

We carried out observations of the $\Lambda$-doubling lines of CH in the $^{2}\Pi_{1/2}, J=1/2$ and the $\Lambda$-doubling lines of OH in the $^{2}\Pi_{3/2}, J=3/2$ with the Arecibo Telescope (Project a2813). We extracted \co\ $J = 1-0$ and \13co\ $J = 1-0$ data from the Five College Radio Astronomy Observatory (FCRAO) Taurus survey~\citep{Taurus_CO}.

\subsection{CH Observations}
\label{CH Observation}

\begin{figure*}[htp]
\includegraphics[width=1.0\linewidth]{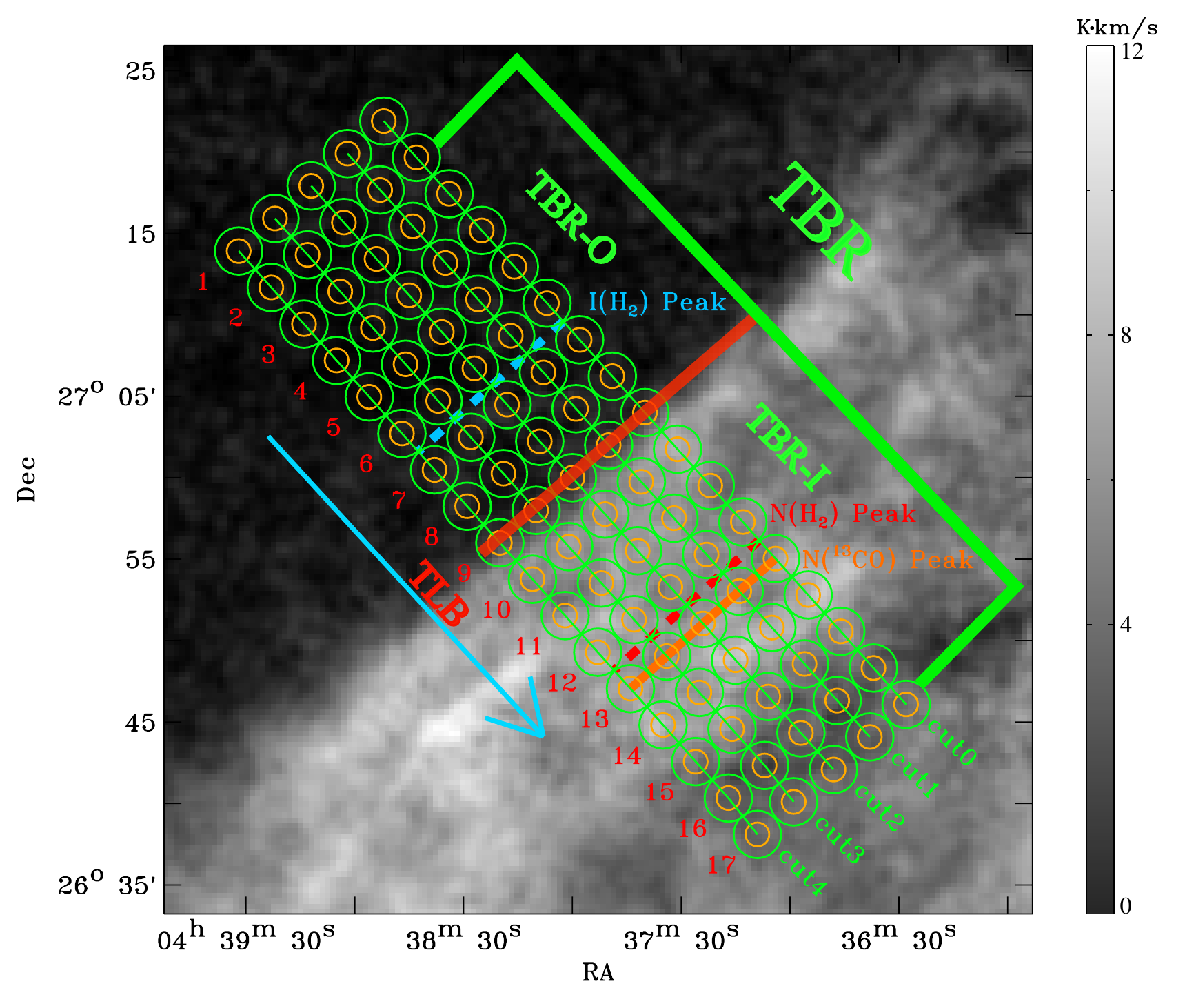}
\caption{Boundary region in \13co\ J = 1-0 peak intensity, with observed positions indicated. The 1.5\arcmin\ yellow circle and the 3\arcmin\ green circle indicate the telescope pointings for CH and OH observation, respectively. The numbers of positions are shown in the figure. The whole Taurus boundary region is denoted as TBR. The Taurus linear boundary (TLB) located at position 9 is shown as a red line. The outside and inside regions of the TBR are abbreviated as TBR-O and TBR-I, respectively. The peak intensity of the two lowest rotational transitions of \h2, S(0) and S(1), is located between position 6 and position 7 \citep{2010ApJ...715.1370G}. The peak column density of \h2\ is located between position 12 and position 13. The peak column density of \13co\ is located at position 13. The arrow in the figure indicates the direction we present spectral line maps. }
\label{fig.cut_ra_dec}
\end{figure*}

The CH observations were taken using the S-high receiver (3.0-4.0 GHz) on 2015 October 25-26 and November 24-26. We observed three $\Lambda$-doubling lines of CH in the $^{2}\Pi_{1/2}, J=1/2$ (as shown in Figure~\ref{fig.ch_energy_level_ground}) at the rest frequencies of 3335.481 (main line $F=1-1$), 3263.794 (lower satellite line $F=0-1$), and 3349.193 MHz (upper satellite line $F=1-0$) with the total power ON mode. Spectra were obtained with the Arecibo WAPP correlator with three-level sampling and 8192 spectral channels for each line in each polarization. The spectral bandwidth was 3.13 MHz for a channel spacing of about 381 Hz, or 0.034 \kms. The average system temperature was about 30 K. The main beam of the antenna pattern had a full width at half maximum (FWHM) beam-width of 1.5\arcmin. Spectra were taken at 17$\times$5 positions across the Taurus boundary region (TBR), as seen in Figure~\ref{fig.cut_ra_dec}. An integration time of 450 s per position was used resulting in RMS noise level of about 0.038 K. To get a higher signal-to-noise ratio, we smoothed the CH spectra to a velocity resolution of 0.14 \kms, which yields {a} RMS noise level of about 0.019 K.

\subsection{OH Observations}
\label{OH Observation}

The OH observations were taken using the L-band wide receiver (1.55-1.82 GHz) on 2013 October 28-31. We observed four $\Lambda$-doubling lines of OH in the $^{2}\Pi_{3/2}, J=3/2$ at the rest frequencies of 1612.231, 1665.402, 1667.359, and 1720.530 MHz with the total power ON mode. Spectra were obtained with the Arecibo WAPP correlator with nine-level sampling and 4096 spectral channels for each line in each polarization. The spectral bandwidth was 3.13 MHz for a channel spacing of about 763 Hz, or 0.142 \kms. The average system temperature was about 31 K. The main beam of the antenna pattern had a FWHM beam-width of 3\arcmin. Spectra were taken at the same positions as those of CH across the TBR, as seen in Figure~\ref{fig.cut_ra_dec}. An integration time of 300 s per position was used resulting in a RMS noise level of about 0.027 K. 

\subsection{$^{12}$CO and $^{13}$CO Data}
\label{co and 13co Data}

The \co\ J = 1-0 and \13co\ J = 1-0 observations were taken simultaneously between  2003 and 2005 using the 13.7 m FCRAO Telescope~\citep{Taurus_CO}. The map is centered at $\alpha(2000.0)=04^h 32^m 44.6^s$, $\delta(2000.0)=24^\circ 25' 13.08''$, with an area of $\sim 98\ \rm deg^2$. The main beam of the antenna pattern has a FWHM beam-width of 45\arcsec\ for \co\ and 47\arcsec\ for \13co. The angular spacing (pixel size) of the resampled on-the-fly data is 20\arcsec~\citep{Goldsmith2008}, which corresponds to a physical scale of $\approx 0.014\rm\ pc$ at a distance of $D=140\ {\rm pc}$. The data have a mean RMS antenna temperature of 0.28 K for \co\ and 0.125 K for \13co. There are 80 and 76 channels with 0.26 and 0.27 \kms\ spacing for \co\ and \13co, respectively.

\section{Analysis}
\label{Analysis}

\subsection{Spectral Analysis}
\label{Spectral Analysis}

\begin{figure*}[htp]
\centering
\includegraphics[width=0.80\linewidth]{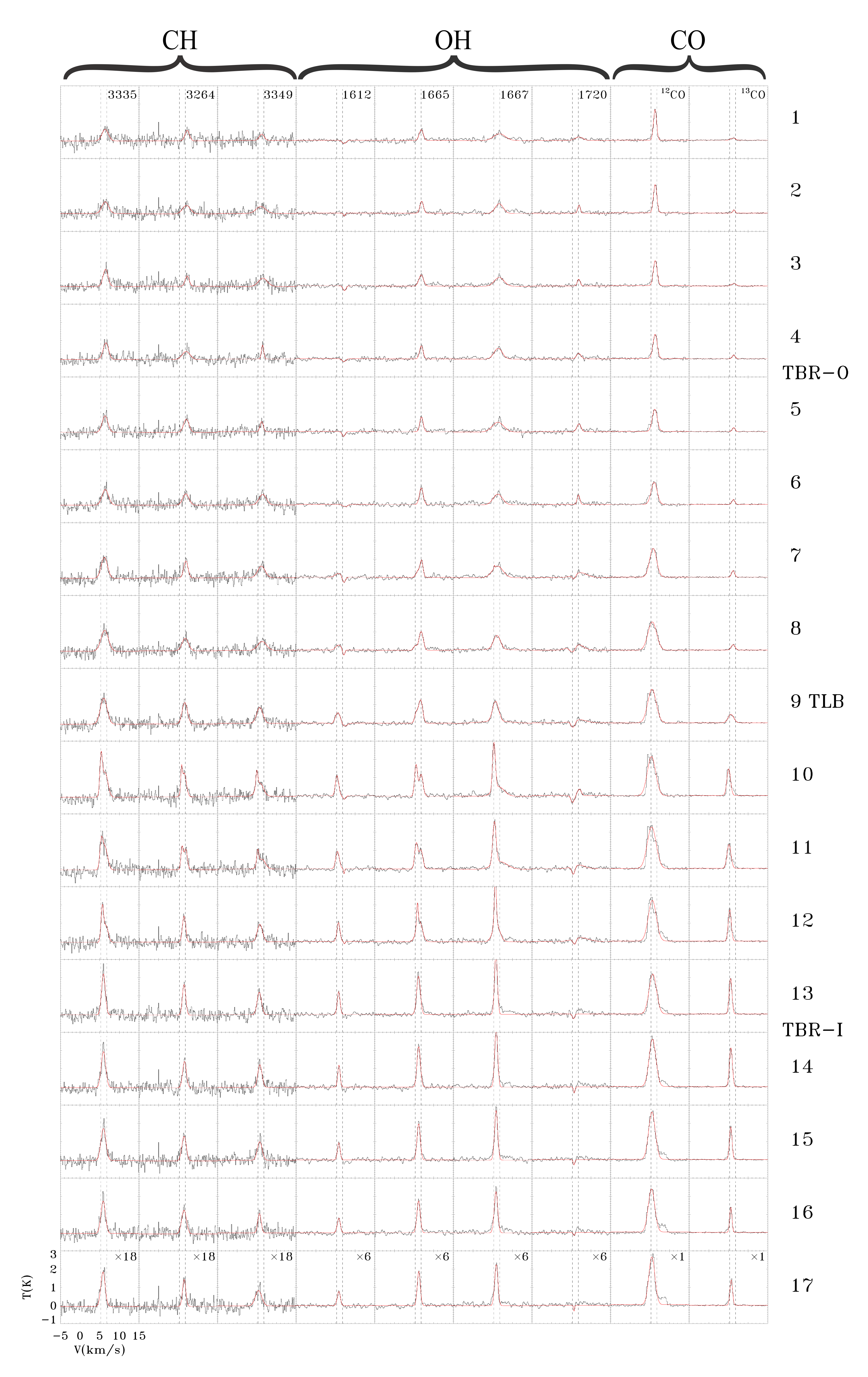}
\caption{Average spectra of all five cuts of CH 3335, 3264, 3349 MHz, OH 1612, 1665, 1667, 1720 MHz, \co\ J = 1-0, and \13co\ J = 1-0 overlaid with corresponding fitted Gaussian profiles (red curve). The \co\ J = 1-0 and \13co\ J = 1-0 spectra were convolved to the OH beam size of 3\arcmin\ at each position. We fitted the CH 3335, 3264, 3349 MHz, OH 1612, 1665, 1667, 1720 MHz spectra with two Gaussian components, and fitted the \co\ J = 1-0 and \13co\ J = 1-0 spectra with single Gaussian component. The vertical dashed lines indicate the central velocities of the two components of OH 1665 MHz at position 10.}
\label{fig.channel_map}
\end{figure*}

\begin{figure*}[htp]
\centering
\includegraphics[width=0.80\linewidth]{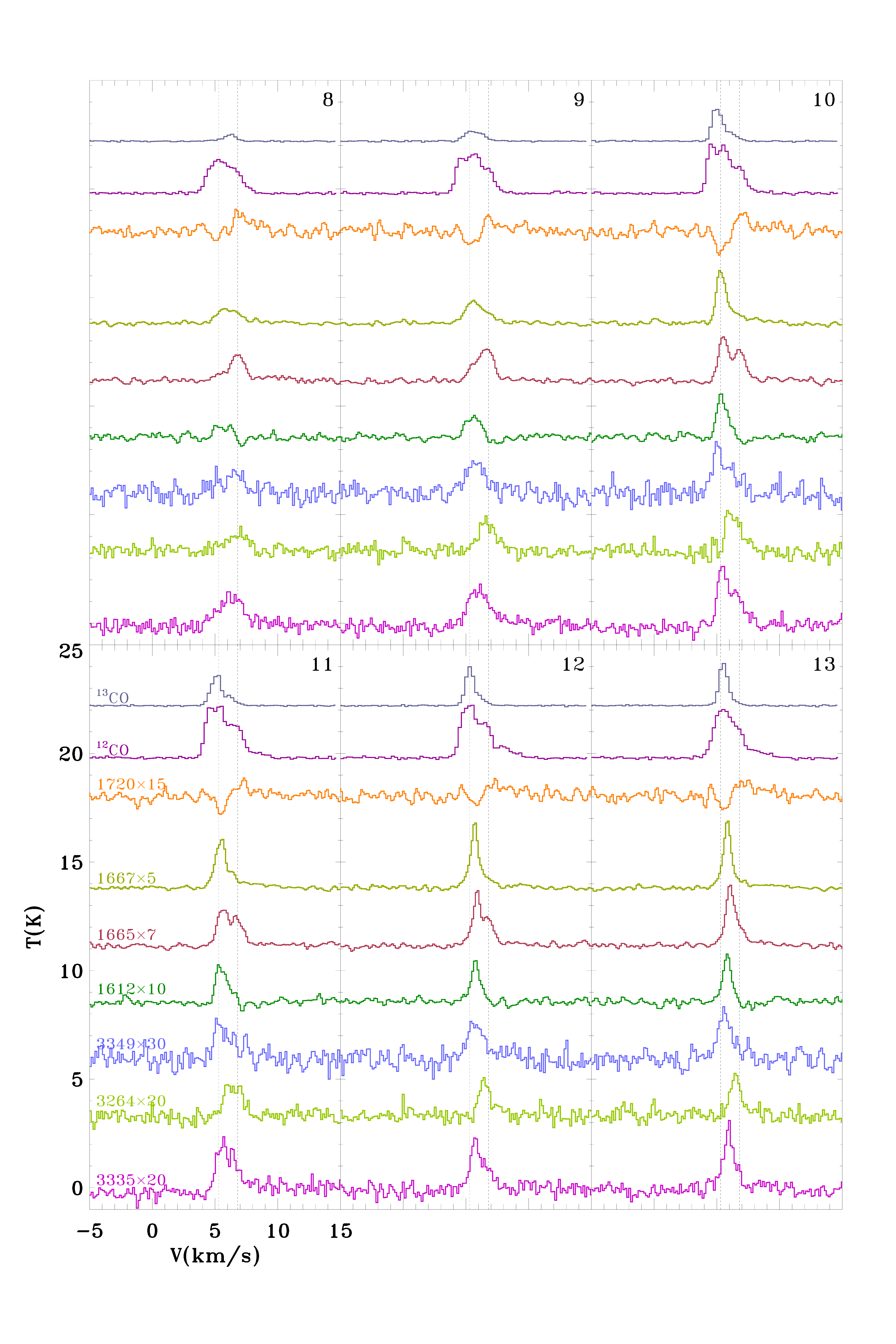}.pdf
\caption{Average spectra of all five cuts of CH 3335, 3264, 3349 MHz, OH 1612, 1665, 1667, 1720 MHz, \co\ J = 1-0, and \13co\ J = 1-0 at positions 8-13. The \co\ J = 1-0, and \13co\ J = 1-0 spectra were convolved to the OH beam size of 3\arcmin\ at each position. The vertical dashed lines indicate the central velocities of the two components of OH 1665 MHz at position 10.}
\label{fig.channel_map_8_13}
\end{figure*}

The locations of the positions for the telescope pointing used to study the TBR are shown in Figure~\ref{fig.cut_ra_dec}. To examine the transition zone with a higher signal-to-noise ratio, we averaged all five cuts of spectra of CH 3335, 3264, 3349 MHz, OH 1612, 1665, 1667, 1720 MHz, \co\ J = 1-0, and \13co\ J = 1-0, as shown in Figure~\ref{fig.channel_map}. The \co\ J = 1-0 and \13co\ J = 1-0 spectra were convolved to the OH beam size of 3\arcmin\ at each position. The emission lines of CH, OH, \co\ J = 1-0, and \13co\ J = 1-0 are well matched in velocity. In particular, the emission lines of CH 3335 MHz and OH 1665 MHz at positions 10--12 all have two components and are well matched in velocity {as shown in Figure~\ref{fig.channel_map_8_13}}.

We fit {a} {two-component} Gaussian to the profiles of the CH 3335, 3264, 3349, OH 1612, 1665, 1667, 1720 MHz spectra and a single Gaussian to \co\ and \13co\ spectra. {The best fitting parameters of CH 3335 MHz, OH 1665 MHz and CO are listed in Table~\ref{tab.ch}-\ref{tab.co}, respectively.} We show the spectra and the fitted profiles in Figure~\ref{fig.channel_map}. \citet{xu2016} have discussed the two components of OH 1665 MHz across the TBR, which indicate the colliding streams or gas flows at the TBR. The CH 3335 MHz spectra also have two components and are well matched with OH 1665 MHz and CO in velocity across the TBR, {as shown in Figure~\ref{fig.vel_all}}, which further confirm the assumption of colliding streams or gas flows across the TBR.

{The shift in the single-component CO velocity is well correlated with the {behavior} of the CH velocity components. When the red (6.5 \kms) CH component is stronger in TBR-O, \13co\ peaks at 6.3 \kms. In contrast, when the blue (5.4 \kms) CH component is stronger in TBR-O, \13co\ peaks at 5.7 \kms. In Figure~\ref{fig.channel_map_8_13} and \ref{fig.vel_all}, the red components of CH and OH gradually become fainter, and disappear at position 13. At the same time, the central velocities of blue components gradually shift from 5.4 \kms\ at position 9 to 5.8 \kms\ at position 13, which indicates that the collision of two streams results in the final central velocity being located between the velocities of the two components. The central velocity of the final combined stream is located closer to the blue components, which have stronger emission lines in TBR-I. This is consistent with the assumption  of different amounts of \13co\ emission at different velocities.}

{The change of the line width of CH {3335 MHz}, OH 1665 MHz, \co~J=1-0 and \13co~J=1-0 along the cut direction is shown in Figure~\ref{fig.width_all}. The line width of the CH 3335 MHz red component in TBR-O is almost a constant {$\sim$1.8 \kms}. After merging with the blue component, the line width slightly decreases to $\sim$1.4 \kms. The line width of OH 1665 MHz remains almost constant across the TBR. The line width of \co, which continues to increase in TBR-O, behaves differently from that of CH 3335 MHz, which is almost constant in TBR-O. 

\citet{1986ApJ...310..392D} pointed out that, owing to the different response of neutral and ionized species to the presence of a magnetic field, the \chp\ and CH formed in the hot post-shock gas can differ significantly in velocities. If shock waves propagate across the TBR, different neutral molecules obtain different velocities owing to their different cross section \citep[e.g.][]{1986ApJ...303..401L, 1993A&A...269..477G}. We find a slightly different central velocities of CH, OH and \co\ across the TBR in Figure~\ref{fig.vel_all}, which may be the result of the propagation of C-shock. More discussion of the probable C-shock across the TBR can be found in Section~\ref{CH Abundance and C-shock Model}. }

{\bf
\begin{table*}[]
\begin{center}
\caption{CH 3335 MHz spectra parameters along the boundary  \label{tab.ch}}
\begin{tabular}{ccccccc}
\hline
Position &\multicolumn{3}{c}{Red Component$^{a}$} & \multicolumn{3}{c}{Blue Component} \\\cline{2-7}\
 ID &   Height$^{b}$ (K) & Center (\kms ) & Width (\kms ) &   Height$^{b}$ (K) & Center (\kms ) & Width (\kms ) \\
 \hline

    1      & - & -  & - & 0.036$\pm$0.004 & 6.5$\pm$0.1  & 1.9$\pm$0.3   \\
    2     & - & -  & - & 0.036$\pm$0.004 & 6.5$\pm$0.1  & 2.0$\pm$0.3    \\
    3     & - & -  & - & 0.052$\pm$0.004 & 6.6$\pm$0.06  & 1.5$\pm$0.1    \\
    4     & - & -  & -  & 0.051$\pm$0.004 & 6.7$\pm$0.06  & 1.6$\pm$0.1   \\
    5     & - & -  & - & 0.050$\pm$0.004 & 6.5$\pm$0.07  & 1.7$\pm$0.2    \\
    6    & - & -  & -  & 0.048$\pm$0.004 & 6.5$\pm$0.07  & 1.9$\pm$0.2    \\
    7      & - & -  & - & 0.064$\pm$0.004 & 6.3$\pm$0.06  & 1.9$\pm$0.1   \\
    8     & - & -  & -  & 0.064$\pm$0.004 & 6.4$\pm$0.06  & 2.2$\pm$0.1   \\
    9     & - & -  & - & 0.080$\pm$0.004 & 6.0$\pm$0.04  & 2.0$\pm$0.1    \\
    10    & 0.12$\pm$0.02 & 5.4$\pm$0.03  & 0.80$\pm$0.1   & 0.075$\pm$0.005 & 6.4$\pm$0.2  & 1.6$\pm$0.3 \\
    11    & 0.071$\pm$0.04 & 5.4$\pm$0.08  & 0.90$\pm$0.3   & 0.070$\pm$0.01 & 6.3$\pm$0.3  & 1.6$\pm$0.5 \\
    12    & 0.10$\pm$0.02 & 5.7$\pm$0.05  & 0.80$\pm$0.1   & 0.050$\pm$0.006 & 6.6$\pm$0.2  & 1.3$\pm$0.4 \\
    13    & 0.13$\pm$0.005 & 6.0$\pm$0.02  & 1.1$\pm$0.1   & - & -  & - \\
    14    & 0.12$\pm$0.005 & 6.0$\pm$0.03  & 1.4$\pm$0.1   & - & -  & - \\
    15    & 0.098$\pm$0.004 & 6.0$\pm$0.03  & 1.5$\pm$0.1   & - & -  & - \\
    16    & 0.098$\pm$0.005 & 5.9$\pm$0.03  & 1.3$\pm$0.1   & - & -  & - \\
    17    & 0.11$\pm$0.005 & 5.8$\pm$0.03  & 1.3$\pm$0.1   & - & -  & - \\
\hline
\multicolumn{7}{p{0.90\textwidth}}{ $^{a}$ After Position 12, Red Component means the converged component.} \\
\multicolumn{7}{p{0.90\textwidth}}{\bf{ $^{b}$ } The height of main beam temperature $T_{MB}$.}\\

\end{tabular}
\end{center}
\end{table*}

\begin{table*}[]
\begin{center}
\caption{OH 1665 MHz spectra parameters along the boundary  \label{tab.oh}}
\begin{tabular}{ccccccc}
\hline
Position &\multicolumn{3}{c}{Red Component$^{a}$} & \multicolumn{3}{c}{Blue Component } \\\cline{2-7}\
 ID &   Height$^{b}$ (K) & Center (\kms ) & Width (\kms ) &   Height$^{b}$ (K) & Center (\kms ) & Width (\kms ) \\
 \hline

    1     & - & -  & -  & 0.10$\pm$0.005 & 6.8$\pm$0.03  & 1.2$\pm$0.1 \\
    2     & - & -  & -  & 0.11$\pm$0.006 & 6.9$\pm$0.03  & 1.1$\pm$0.1 \\
    3     & - & -  & -  & 0.10$\pm$0.005 & 6.8$\pm$0.03  & 1.4$\pm$0.1 \\
    4     & - & -  & -  & 0.12$\pm$0.006 & 6.9$\pm$0.03  & 1.1$\pm$0.1 \\
    5     & - & -  & -  & 0.13$\pm$0.006 & 6.9$\pm$0.02  & 1.0$\pm$0.1 \\
    6     & - & -  & -  & 0.15$\pm$0.006 & 6.9$\pm$0.02  & 1.2$\pm$0.1 \\
    7     & 0.06$\pm$0.01 & 6.1$\pm$0.3  & 1.2$\pm$0.4  & 0.14$\pm$0.02 & 6.9$\pm$0.07  & 0.8$\pm$0.1 \\
    8     & 0.04$\pm$0.01 & 5.3$\pm$0.09  & 0.9$\pm$0.2  & 0.17$\pm$0.01 & 6.8$\pm$0.02  & 1.3$\pm$0.1 \\
    9     & 0.09$\pm$0.01 & 5.5$\pm$0.08  & 1.0$\pm$0.1  & 0.21$\pm$0.01 & 6.7$\pm$0.04  & 1.2$\pm$0.1 \\
    10    & 0.29$\pm$0.01 & 5.5$\pm$0.01  & 0.9$\pm$0.03  & 0.20$\pm$0.01 & 6.8$\pm$0.02  & 1.1$\pm$0.1 \\
    11    & 0.23$\pm$0.01 & 5.6$\pm$0.03  & 1.0$\pm$0.05  & 0.17$\pm$0.01 & 6.8$\pm$0.04  & 1.2$\pm$0.1 \\
    12    & 0.20$\pm$0.01 & 5.9$\pm$0.01  & 0.5$\pm$0.03  & 0.19$\pm$0.01 & 6.3$\pm$0.03  & 1.8$\pm$0.1 \\
    13    & 0.35$\pm$0.01 & 6.1$\pm$0.01  & 1.1$\pm$0.02  & - & -  & - \\
    14    & 0.37$\pm$0.01 & 6.2$\pm$0.01  & 1.0$\pm$0.02  & - & -  & - \\
    15    & 0.33$\pm$0.01 & 6.2$\pm$0.01  & 1.0$\pm$0.02  & - & -  & - \\
    16    & 0.29$\pm$0.01 & 6.2$\pm$0.01  & 1.0$\pm$0.02  & - & -  & - \\
    17    & 0.31$\pm$0.01 & 6.2$\pm$0.01  & 0.9$\pm$0.02  & - & -  & - \\
\hline
\multicolumn{7}{p{0.90\textwidth}}{ $^{a}$ After Position 12, Red Component means the converged component.} \\
\multicolumn{7}{p{0.90\textwidth}}{\bf{ $^{b}$ } The height of main beam temperature $T_{MB}$.}\\

\end{tabular}
\end{center}
\end{table*}

\begin{table*}[]
\begin{center}
\caption{CO spectra parameters along the boundary  \label{tab.co}}
\begin{tabular}{ccccccc}
\hline
Position &\multicolumn{3}{c}{\co} & \multicolumn{3}{c}{\13co} \\\cline{2-7}\
 ID &   Height$^{a}$ (K) & Center (\kms ) & Width (\kms ) &   Height$^{a}$ (K) & Center (\kms ) & Width (\kms ) \\
 \hline

    1     & 1.7$\pm$0.04  & 6.3$\pm$0.01  & 0.9$\pm$0.03  & 0.1$\pm$0.01  & 6.2$\pm$0.07  & 1.4$\pm$0.17 \\
    2     & 1.6$\pm$0.04  & 6.4$\pm$0.01  & 1.0$\pm$0.03  & 0.1$\pm$0.02  & 6.3$\pm$0.06  & 1.0$\pm$0.14 \\
    3     & 1.4$\pm$0.04  & 6.4$\pm$0.01  & 1.1$\pm$0.03  & 0.1$\pm$0.01  & 6.4$\pm$0.07  & 1.4$\pm$0.17 \\
    4     & 1.3$\pm$0.03  & 6.3$\pm$0.01  & 1.3$\pm$0.04  & 0.2$\pm$0.01  & 6.3$\pm$0.02  & 0.9$\pm$0.06 \\
    5     & 1.2$\pm$0.04  & 6.2$\pm$0.02  & 1.5$\pm$0.06  & 0.2$\pm$0.01  & 6.3$\pm$0.02  & 0.9$\pm$0.06 \\
    6     & 1.2$\pm$0.04  & 6.1$\pm$0.03  & 1.8$\pm$0.07  & 0.3$\pm$0.01  & 6.3$\pm$0.02  & 0.9$\pm$0.05 \\
    7     & 1.6$\pm$0.03  & 5.8$\pm$0.02  & 2.1$\pm$0.05  & 0.4$\pm$0.01  & 6.2$\pm$0.02  & 0.9$\pm$0.04 \\
    8     & 1.6$\pm$0.03  & 5.6$\pm$0.03  & 2.6$\pm$0.07  & 0.3$\pm$0.01  & 6.2$\pm$0.02  & 1.1$\pm$0.06 \\
    9     & 1.8$\pm$0.05  & 5.5$\pm$0.03  & 2.5$\pm$0.08  & 0.5$\pm$0.01  & 5.6$\pm$0.02  & 1.7$\pm$0.06 \\
    10    & 2.2$\pm$0.08  & 5.4$\pm$0.04  & 2.5$\pm$0.11  & 1.5$\pm$0.05  & 5.1$\pm$0.02  & 1.1$\pm$0.04 \\
    11    & 2.3$\pm$0.08  & 5.5$\pm$0.05  & 2.7$\pm$0.12  & 1.3$\pm$0.04  & 5.1$\pm$0.02  & 1.2$\pm$0.05 \\
    12    & 2.2$\pm$0.08  & 5.6$\pm$0.04  & 2.4$\pm$0.11  & 1.7$\pm$0.04  & 5.3$\pm$0.01  & 0.9$\pm$0.03 \\
    13    & 2.2$\pm$0.05  & 5.8$\pm$0.03  & 2.2$\pm$0.07  & 1.9$\pm$0.03  & 5.5$\pm$0.01  & 0.8$\pm$0.02 \\
    14    & 2.6$\pm$0.05  & 5.7$\pm$0.02  & 2.0$\pm$0.05  & 2.1$\pm$0.03  & 5.6$\pm$0.01  & 0.9$\pm$0.01 \\
    15    & 2.6$\pm$0.06  & 5.5$\pm$0.02  & 2.0$\pm$0.06  & 1.8$\pm$0.02  & 5.6$\pm$0.00  & 0.8$\pm$0.01 \\
    16    & 2.4$\pm$0.07  & 5.4$\pm$0.03  & 2.0$\pm$0.07  & 1.4$\pm$0.03  & 5.6$\pm$0.01  & 0.7$\pm$0.02 \\
    17    & 2.6$\pm$0.08  & 5.5$\pm$0.03  & 1.8$\pm$0.07  & 1.4$\pm$0.04  & 5.7$\pm$0.01  & 0.8$\pm$0.02 \\
\hline
\multicolumn{7}{p{0.90\textwidth}}{\bf{ $^{a}$ } The height of main beam temperature $T_{MB}$.}\\

\end{tabular}
\end{center}
\end{table*}

\begin{figure}[htp]
\centering
\includegraphics[width=1.0\linewidth]{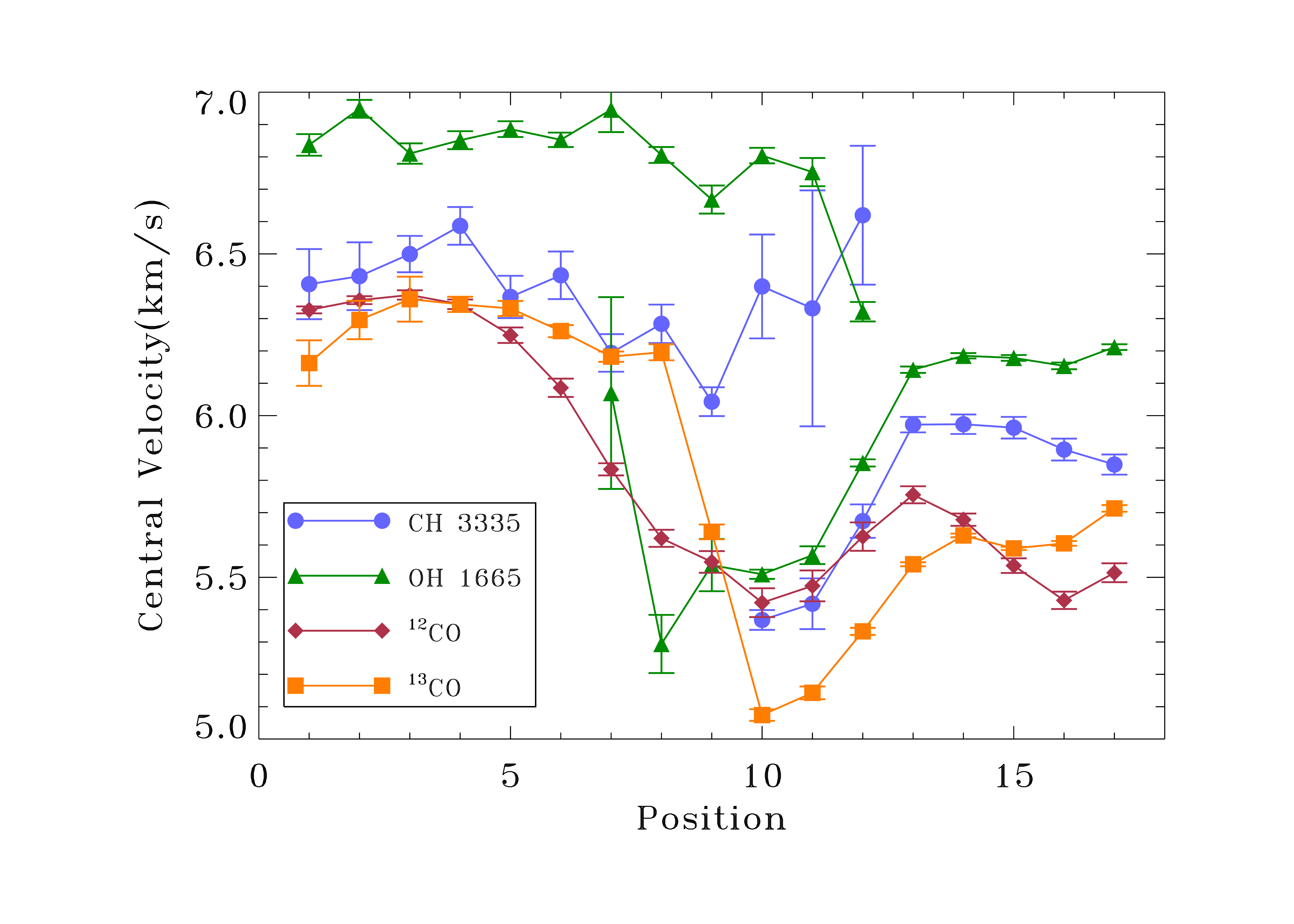}
\caption{The change of central velocities of CH {3335 MHz}, OH 1665 MHz, \co~J=1-0 and \13co~J=1-0 along the cut direction shown in Figure~\ref{fig.cut_ra_dec}.}
\label{fig.vel_all}
\end{figure}

\begin{figure}[htp]
\centering
\includegraphics[width=1.0\linewidth]{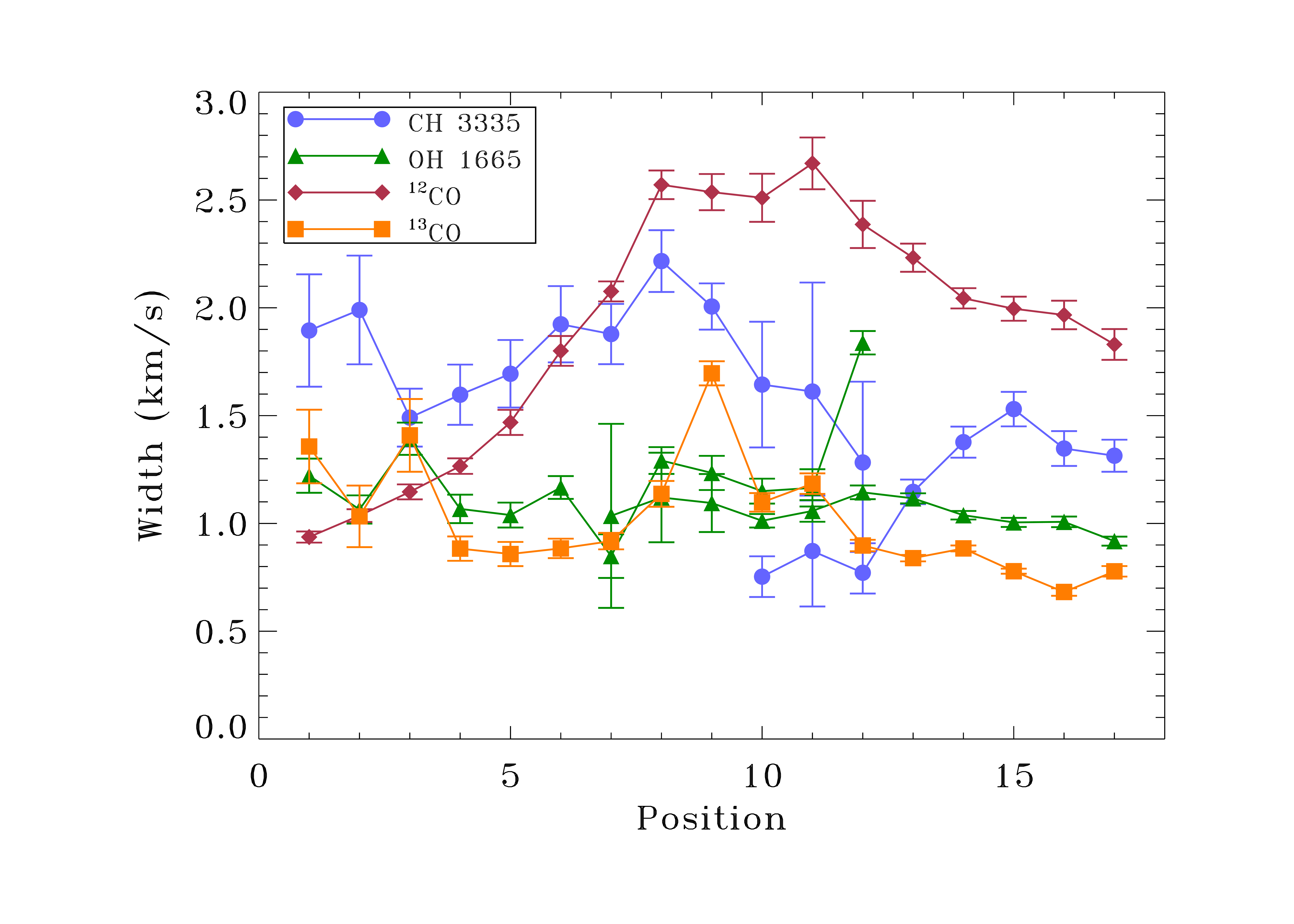}
\caption{The change of line width of CH {3335 MHz}, OH 1665 MHz, \co~J=1-0 and \13co~J=1-0 along the cut direction shown in Figure~\ref{fig.cut_ra_dec}.}
\label{fig.width_all}
\end{figure}

}

\subsection{CH Column Density}
\label{CH Column Density}

The TBR has a relatively low UV field between $\chi=0.3$ and 0.8 in units of the Draine's field \citep{2009ApJ...701.1450F, pineda2010}. The gas kinetic temperature does not exceed 40 K even in the outermost layer of the TBR, and can be as low as 10 K in the inner part of the TBR. This means that almost all the CH molecules are populated in the $^{2}\Pi_{1/2}, J=1/2$ $\Lambda$-doubling levels (as shown in Figure~\ref{fig.ch_energy_level_ground}), and the populations of the higher $J$ levels are negligible. Although the gas kinetic temperature is approximately 40 K in the outermost layer of the TBR due to the heating by interstellar UV radiation, the density there is too low ($\sim60$ \cm3) to excite CH even to the first rotationally excited state ($J = 3/2$). The critical density required to excite the lowest rotational transition ($J = 3/2-1/2$, 530 GHz) is as high as $10^{6}$ \cm3 \citep{2012A&A...546A.103S}. Hence, practically all CH molecules are in the $J = 1/2$ levels. For this reason, we only considered the $J = 1/2$ levels for calculations of the partition function in the calculation of CH column density. 

In most of the observation, the populations of the $^{2}\Pi_{1/2}, J=1/2$ $\Lambda$-doubling levels can be inverted over a wide range of physical conditions causing weak masers producing negative excitation temperatures. In the case of the main line, the excitation temperature $T_{\rm ex,11}$ for CH is in the range $\sim$ -60 to -10 K \citep{1977ApJS...35..263H, 1979A&A....73..253G, 1984ApJ...285..312B, 1988ApJ...329..920S, 2002A&A...391..693L}.

We calculated the column density of CH assuming it to be optically thin, which is reasonable owing to the very small Einstein A-coefficients  of the transitions $A=1.94\times10^{-10}$ $\rm s^{-1}$:
\begin{equation}
N({\rm CH})=2.82\times10^{14}\frac{1}{1-T_{\rm bg}/T_{\rm ex,11}}\int{}{}T_{\rm MB}(3335){\rm d}v \lc
\end{equation}
where $T_{\rm MB}(3335)$ is the main-beam brightness temperature of the main $\Lambda$-doubling line ($F=1-1$) in $^{2}\Pi_{1/2}$, $T_{\rm bg}$ is the cosmic background temperature, and $T_{\rm ex,11}$ is the excitation temperature of the $F=1-1$ transition. 

\begin{figure}[htp]
\centering
\includegraphics[width=1.0\linewidth]{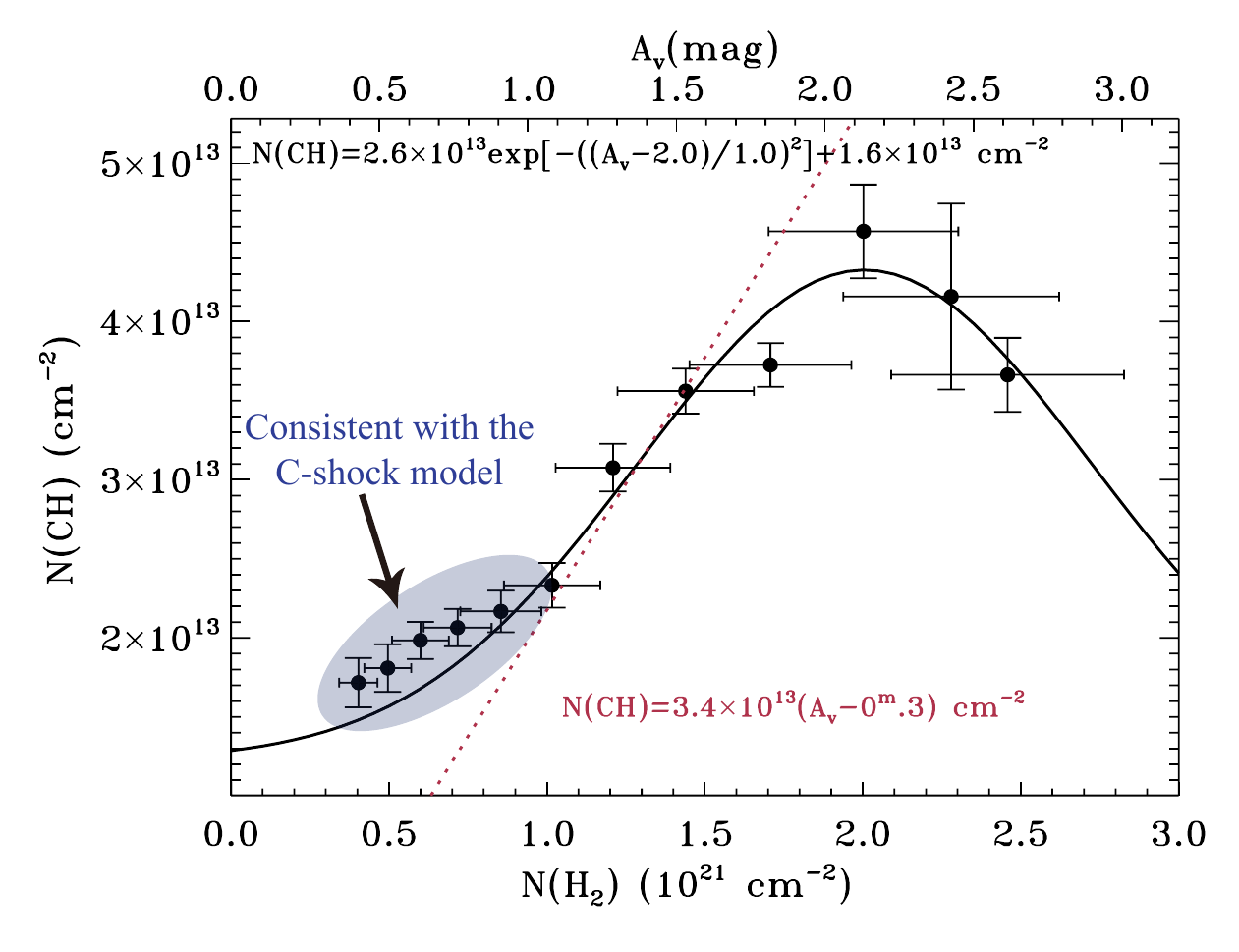}
\caption{Change of column density of CH across the TBR as a function of visual extinction $A_{v}$ and total gas column density. The dashed line indicates the observational relation between the CH column density and visual extinction for diffuse, dark and molecular cloud \citep{1986A&A...160..157M}. The red shade indicates the overabundance of CH below 1 mag extinction. This overabundance is consistent with the C-shock model prediction.}
\label{fig.ch_col}
\end{figure}

We have to choose a value for $T_{\rm ex,11}$ to calculate the CH column density. As mentioned above, in most of the observation, CH emission can be weak masers yielding to negative excitation temperatures $T_{\rm ex,11}$ in the range $\sim-60$ to $-10$ K. We adopted $T_{\rm ex,11}=-15$ K which is in the range of previously observed temperatures and also has also been widely used in diffused gas \citep{1978ApJ...224..125L, 2002A&A...391..693L}. When $T_{\rm ex,11}=-15$ K, the correction factor for excitation temperature,
\begin{equation}
f_{\rm ex,bg}=\frac{1}{1-T_{bg}/T_{\rm ex,11}}\lc
\end{equation}
is about 0.85. We parametrize the trend of CH column density across the TBR in a Gaussian profile, as shown in Figure~\ref{fig.ch_col}. The trend of CH column density can be well described as
 \begin{equation}
 \label{ch_col_eq}
{N({\rm{CH}})}=2.6\times10^{13}\exp[-(\frac{A_{v}-2.0}{1.0})^2]+1.6\times10^{13}\  {\rm{cm^{-2}}}\lp
\end{equation}
When extinction exceeds 2 mag in TBR, CH column density starts to drop, indicating the consumption of CH via the carbon chemistry in relatively dense molecular environments, with more carbon locked into CO \citep{1986A&AS...64..391V}.

When using the value of $T_{\rm ex,11}=-60$ K from \citet{1979A&A....73..253G}, which is widely used for dark clouds, $f_{\rm ex,bg}$ increases by about 11\%. The uncertainty in CH column density associated with the assumptions of $T_{\rm ex,11}$ is thus small.

The assumption that the $\Lambda$-doubling lines of CH are optically thin can be problematic, which can be seen from the relative intensities of the three components (Figure~\ref{fig.channel_map}). Both satellite lines should be two times weaker than the main component, but the observed intensities of satellite lines are often larger. Furthermore, the intensity of the two observed satellite lines are not equal. Thus, we cannot precisely calculate the optical depth of CH. The correction factor for optical depth is defined as
\begin{equation}
f_{\tau}=\frac{\tau}{1-e^{-\tau}}\lc
\end{equation}
where $\tau$ is the optical depth of the main $\Lambda$-doubling line. The line ratio between the satellite line and the main line ranges from 0.5 to 0.9, resulting in $1\le f_{\tau} \le 2.6$. Considering all the factors above (including $T_{\rm ex,11}$ and $\tau$), the CH column density can be underestimated by as much as a factor of 3.

\section{CH As a Molecular Gas Tracer}
\label{CH As a Molecular Gas Tracer}

\begin{figure*}[htp]
\centering
\includegraphics[width=0.9\linewidth]{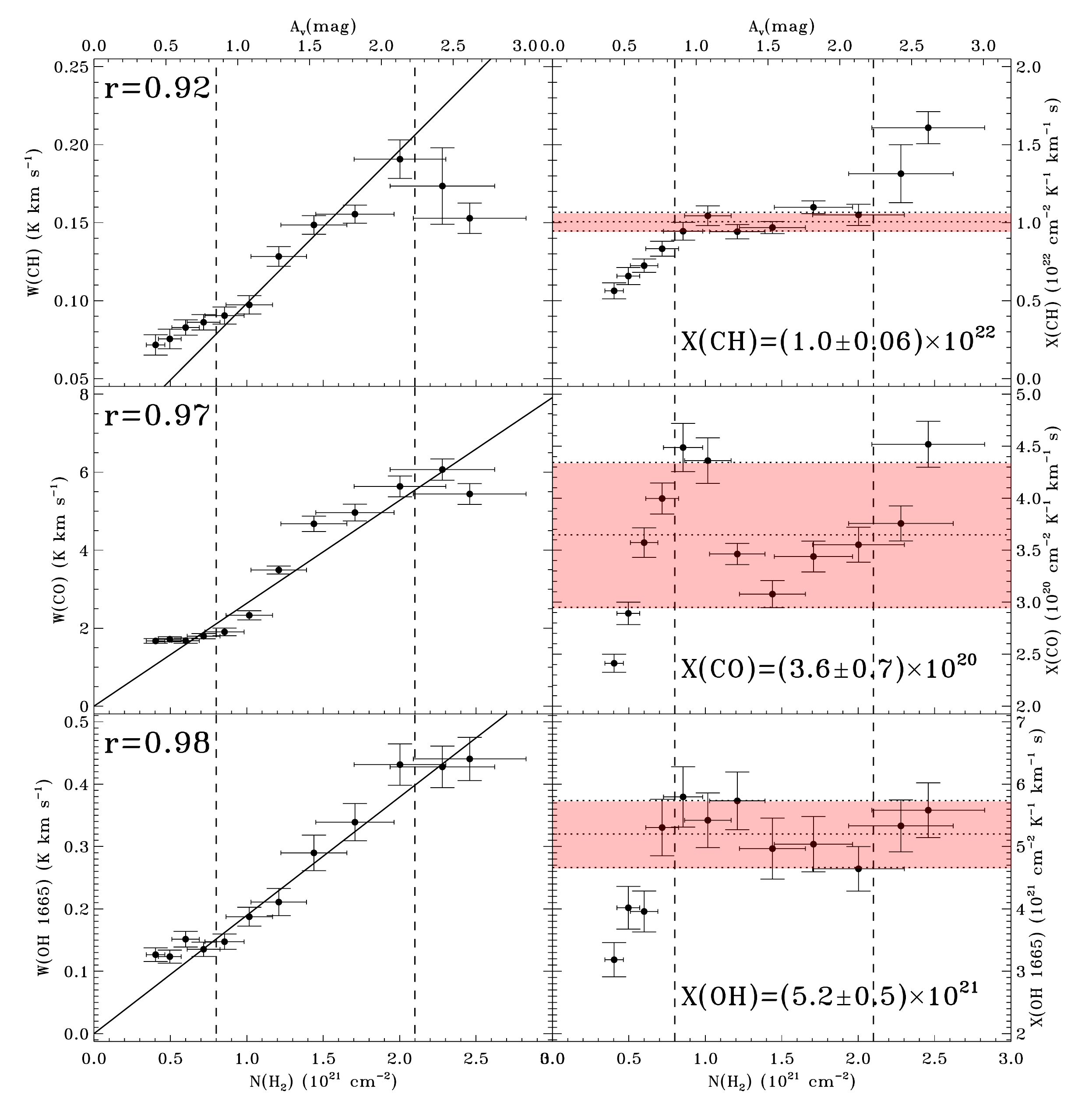}
\caption{Correlation plot of the integrated intensity $W$ of CH, OH and CO versus the visual extinction $A_{v}$, and the correlation of the ``$X$-factor'' of CH, OH and CO versus the visual extinction $A_{v}$. ``r'' means Pearson correlation coefficient. }
\label{fig.av_x}
\end{figure*}

Based on dust extinction and total gas column density provided by previous studies \citep{pineda2010, 2014ApJ...795...26O, xu2016}, we examine the evolution of CH across TBR in terms of $X$-factor. We show the correlation of the integrated intensity $W$ (in K \kms) of CH 3335 MHz, OH 1665 MHz and \co\ 1-0 versus the visual extinction in Figure~\ref{fig.av_x}. Over a wide visual extinction range (0.4-2.7 mag), W(CO) and W(OH) correlate better with $A_{v}$ than W(CH) does. But in a more limited extinction range between 0.8 and 2.1 mag, more appropriate for a translucent cloud and/or the transition zone of a dark cloud, there is a better correlation between W(CH) and $A_{v}$ than those between W(OH), W(CO) and $A_{v}$. To better examine these relationships, we adopt the usual definition of  ``$X$-factor'',
\begin{equation}
X{\rm-factor}=N_{\rm H2}/W \lc
\end{equation}
where $N_{\rm H2}$ is the column density of \h2\ and W is the integrated intensity of the molecular tracers CH, OH, CO. We plot the correlation of ``$X$-factor'' of CH, OH and CO versus the visual extinction in Figure~\ref{fig.av_x}. When the visual extinction $A_{v}$ is in the range between 0.8 and 2.1 mag, the $X_{\rm CH}$-factor is almost a constant of $(1.0\pm0.06)\times10^{22}$ $\rm {cm^{-2}~K^{-1}~km^{-1}~s}$. The dispersion of CH $X$-factor is visibly less than those of OH and CO at $0.8\le A_{v} \le 2.1$ mag. CH appears to be a better tracer of molecular gas than CO and OH in the transition zone ($0.8\le A_{v} \le 2.1$ mag). Where extinction drops below 1 mag in TBR, the integrated intensities W of the three molecules lie above the overall fitting lines, indicating a stronger intensity of these spectra, which is likely the result of C-shock in the TBR-O (Section~\ref{CH Abundance and C-shock Model}). 
 
\section{CH Abundance and C-Shock Model}
\label{CH Abundance and C-shock Model}

We have calculated the column density of CH in Section~\ref{CH Column Density}. The column density of CH across the TBR is shown in Figure~\ref{fig.ch_col}. \citet{1986A&A...160..157M} found a good correlation between the CH column density and visual extinction for diffuse, dark molecular cloud $N({\rm CH})=3.4\times10^{13}(A_{v}-0^{m}.3)~{\rm cm^{-2}}$ as the dashed line shows in Figure~\ref{fig.ch_col}. This correlation is reasonable considering the CH formation threshold, which requires substantial \h2\ \citep{1973ApL....15...79B}. The derived column density of CH below 1 mag extinction ({blue} shade in Figure~\ref{fig.ch_col}) in TBR obviously lies above the typical value observed by \citet{1986A&A...160..157M}. Furthermore, considering the correction factor for excitation temperature and optical depth (as discussed in Section~\ref{CH Column Density}), the actual column density of CH in TBR-O is likely to be even larger by an additional factor of 3. The true CH column density in TBR-O can thus be one order of magnitude higher than the simple extension {from} the $X$-factor for higher extinctions.

We compared the CH column density with the prediction of C-shock models. \citet{1998MNRAS.297.1182F} made a prediction that CH column density would increase from $10^{13}$ to $10^{14}$ \2cm\ when extinction increases from 0.1 to 1 mag at a C-shock front{, which matches exactly the measured values here.} \citet{xu2016} also found an overabundance of OH below 1 mag extinction (by a factor of 80). The overabundance of CH and OH suggests that there may be an additional channel of CH and OH production, possibly due to the shock \citep[e.g.][]{1986ApJ...306..655D} produced by the colliding streams \citep[][Figure~4]{xu2016}. When shock waves propagate through the molecular ISM, the gas is compressed, heated, and accelerated. CH can be produced during \chp\ synthesis (Equation~(\ref{chreact2})) in lower density material ($\sim 50$ \cm3) from MHD shocks \citep{1986ApJ...306..655D, 1986MNRAS.220..801P}. In a related work, \citet{2010ApJ...715.1370G} found anomalous {rotationally} excited \h2, indicating high gas temperature exceeding 200 K in TBR-O. Such high temperatures cannot be reproduced in PDR models \citep{2010ApJ...715.1370G} {while being consistent with} the existence of shocks. {\citet{2014ApJ...795...26O} also ruled out PDR models due to non-detection of [\cii] in TRR-O.} If temperature is above 300 K, the neutral-neutral reactions become important, which can also result in the overabundance of OH \citep{2002ApJ...580..278N, xu2016}.

{We compared the CH column density with UV-driven PDR models \citep[e.g.][]{2012A&A...544A..22L, 2013A&A...550A..56R, 2016ApJ...829...15M}. \citet{2016ApJ...829...15M} found that a steady-state UV-driven PDR chemistry with radiation field $\chi=(1-5)\times 10^{3}$ (in Draine units), rather than a shock chemistry, plays a key role in \chp\ formation in Orion BN/KL. However, the TBR has a relatively low UV field between $\chi=0.3$ and 0.8 \citep{2009ApJ...701.1450F, pineda2010}. {When the radiation field strength declines}, the number density of \chp\ decreases around $A_{v}=1$ mag \citep{1992MNRAS.255..463D}. \citet{2012A&A...544A..22L} and \citet{2013A&A...550A..56R} made a PDR model with {a} relatively low radiation field with $\chi=1-10$. Neither model can form as much CH as in our {observations}. In particular, \citet{2012A&A...544A..22L} made predictions of CH column density in a UV-driven chemistry simulation with radiation field of $\chi=1$. The observed CH column density is still 2-3 times lager than their predictions. As discussed in Section~\ref{CH Column Density}, the actual column density of CH in TBR-O is likely to still be larger by another factor of 3. The observed CH column density can thus be as much as 6-9 times larger than that of UV-driven simulation. UV photons should play only a limited role in producing CH in a relatively low radiation field such as in the TBR. 

Besides the C-shock model and UV-driven PDR model, \citet{1993A&A...269..477G} and \citet{1995MNRAS.277..458C} observed optical CH and \chp\ lines in several stars but found their line profiles are inconsistent with shock theories, and predict a significant velocity difference between neutral and ionized species, as discussed in Section~\ref{Spectral Analysis}. They believed that turbulent chemistry may play a key role in the overabundant of CH and \chp\ in these clouds. Since we do not have ionized species data (e.g. \chp ), we cannot examine the proposed velocity difference between neutral and ionized species. In terms of neutral species, CH, OH, and \co\ have mildly different central velocities (Figure~\ref{fig.vel_all}), consistent with the propagation of C-shock. Most turbulent dissipation region models \citep[e.g.][]{2014A&A...570A..27G} predict two orders of magnitude more \chp\ than the PDR model, which could explain the overabundance of CH in TBR-O. If the region is turbulence dominated, the line width of different molecules is likely to have a linear relation. But in Figure ~\ref{fig.width_all}, the line width of CH, OH, \co\ and \13co\ dose not show clear correlation in TBR-O, which does not support turbulence dissipation. 
}

\section{Summary and Conclusions}
\label{Summary and Conclusions}

We have mapped a sharp boundary region of the Taurus molecular cloud in all three ground-state transitions of the methylidyne (CH) radical with the Arecibo telescope. A combined analysis of CH data with OH, \co\ J = 1-0, \13co\ J = 1-0{,} and dust leads to the following conclusions:

\begin{enumerate}[1.]
\item CH has two kinematic components. One component shifts from 5.3 to 6 \kms\ going from outside to inside, {both of which match well with those of OH.} The shifting of the two kinematic components indicates colliding streams or gas flow at the boundary region.

\item The derived CH abundance across the boundary is consistent with the previous observation for $0.8\le A_{v} \le 2.1$ mag, but overabundant by as much as one order of magnitude below 1 mag extinction. The overabundance of CH is consistent with the prediction of the C-shock model {rather than a UV-driven PDR model}, which supports the existence of C-shock across the TBR. CH can be produced during \chp\ synthesis in low density material from C-shocks. 

\item The scatter of CH $X$-factor ($1.0\pm0.06)\times10^{22}$ is much smaller than those of CO and OH in the transition zone ($0.8\le A_{v} \le 2.1$ mag). CH is thus potentially a better tracer of molecular gas than CO or OH for translucent clouds, cloud boundaries, and the transition zone.

\end{enumerate}

\acknowledgments
\begin{acknowledgements}
This work is partly supported by the China Ministry of Science and Technology under State Key Development Program for Basic Research (973 program) No. 2012CB821802, the National Natural Science Foundation of China No. 11373038, No. 11373045, and the Strategic Priority Research Program "The Emergence of Cosmological Structures" of the Chinese Academy of Sciences, Grant No. XDB09010302. 

We are grateful to Paul Goldsmith for his kind and valuable advice. {We would like to thank the anonymous referee for the careful inspection of the manuscript and constructive comments, particularly the important suggestion to add the comparison with the UV-driven PDR model to improve the quality of this study.}

\end{acknowledgements}

\end{CJK*}

\end{document}